\documentclass[12pt, epsfig]{article}                            
\usepackage{epsfig}       
\usepackage{amssymb}     
\usepackage{amsfonts} 
\usepackage[square]{natbib}
\usepackage{graphicx}
\usepackage{dcolumn}
\usepackage{bm}
\usepackage{amsmath}
\usepackage{amsxtra}
\usepackage{amstext}
\usepackage{latexsym}
 \newskip\humongous \humongous=0pt plus 1000pt minus 1000pt

\newif\ifdtup

\catcode`\@=11
 
\@addtoreset{equation}{section}
\def\theequation{\thesection.\arabic{equation}}
 
\def\@normalsize{\@setsize\normalsize{15pt}\xiipt\@xiipt
\abovedisplayskip 14pt plus3pt minus3pt%
\belowdisplayskip \abovedisplayskip
\abovedisplayshortskip \z@ plus3pt%
\belowdisplayshortskip 7pt plus3.5pt minus0pt}
 
\def\small{\@setsize\small{13.6pt}\xipt\@xipt
\abovedisplayskip 13pt plus3pt minus3pt%
\belowdisplayskip \abovedisplayskip
\abovedisplayshortskip \z@ plus3pt%
\belowdisplayshortskip 7pt plus3.5pt minus0pt
\def\@listi{\parsep 4.5pt plus 2pt minus 1pt
     \itemsep \parsep
     \topsep 9pt plus 3pt minus 3pt}}
 
\relax

\catcode`@=12
 
\catcode`\@=11
 
\def\section{\@startsection{section}{1}{\z@}{3.5ex plus 1ex minus
   .2ex}{2.3ex plus .2ex}{\large\bf}}
 
\def\thesection{\arabic{section}}    
\def\thesubsection{\arabic{section}.\arabic{subsection}}

\def\appendix{\setcounter{section}{0}
 \def\thesection{Appendix \Alph{section}}
 \def\thesubsection{\Alph{section}.\arabic{subsection}}
 \def\theequation{\Alph{section}.\arabic{equation}}}

\begin{document}

\newcommand{\beq}{\begin{equation}}
\newcommand{\eeq}{\end{equation}}
\newcommand{\bea}{\begin{eqnarray}}
\newcommand{\eea}{\end{eqnarray}}
\newcommand{\beas}{\begin{eqnarray*}}
\newcommand{\eeas}{\end{eqnarray*}}
\newcommand{\defi}{\stackrel{\rm def}{=}}
\newcommand{\non}{\nonumber}
\newcommand{\bquo}{\begin{quote}}
\newcommand{\enqu}{\end{quote}}
\def\de{\partial}
\def\Tr{ \hbox{\rm Tr}}
\def\const{\hbox {\rm const.}}
\def\o{\over}
\def\im{\hbox{\rm Im}}
\def\re{\hbox{\rm Re}}
\def\bra{\langle}\def\ket{\rangle}
\def\Arg{\hbox {\rm Arg}}
\def\Re{\hbox {\rm Re}}
\def\Im{\hbox {\rm Im}}
\def\diag{\hbox{\rm diag}}
\def\longvert{{\rule[-2mm]{0.1mm}{7mm}}\,}
\def\a{\alpha}
\def\dag{{}^{\dagger}}
\def\tq{{\widetilde q}}
\def\p{{}^{\prime}}
\def\W{{\cal W}}
\def\th{\theta}
\def\b{\beta}
\def\t{\tau}
\def\mj{\mathcal{J}}
\newcommand{\Z}{\ensuremath{\mathbb Z}}
\renewcommand{\,}{,\hspace{.5cm}}
\newcommand{\ddx}{\frac{\partial}{\partial x}}
\newcommand{\ddt}{\frac{\partial}{\partial \tau}}
\newcommand{\pp}{{}^{\prime\prime}}
\newcommand{\hsp}{,\hspace{.5cm}}
\newcommand{\gt}{\tilde{\gamma}}
\newcommand{\vt}{\tilde{V}}
\newcommand{\se}{\hspace{-.1cm}=\hspace{-.1cm}}
\renewcommand{\sp}{\hspace{-.1cm}+\hspace{-.1cm}}

\begin{titlepage}
\begin{flushright}
ULB-TH/05-27\\
\end{flushright}
\def\thefootnote{\fnsymbol{footnote}}

\begin{center}
{\large {\bf Geometrical Thermodynamic Field Theory
 } }
\end{center}
\begin{center}
{\large  Giorgio Sonnino\footnote{\texttt{gsonnino@ulb.ac.be}} {\it and} Jarah Evslin\footnote{\texttt{jevslin@ulb.ac.be}} }
\end{center}

\begin{center}
{\it 
EURATOM - Belgian State Fusion Association 

and 

International Solvay Institutes 

Department of Theoretical Physics and Mathematics

Free University of Brussels (U.L.B.), Blvd du Triomphe 

Campus de la Plaine, C.P. 231, Building  NO}

Brussels, B-1050, Belgium\\

\end{center}

\vspace{.5cm}

\begin{abstract}
A manifestly covariant, coordinate independent reformulation of the Thermodynamic Field Theory (TFT) is presented. The TFT is a covariant field theory that describes the evolution of a thermodynamic system, extending the near-equilibrium theory established by Prigogine in 1954.  We introduce the {\it Minimum Dissipation Principle}, which is conjectured to apply to any system relaxing towards a steady-state. We also derive the thermodynamic field equations, which in the case of $\alpha-\alpha$ and $\beta-\beta$ processes have already appeared in the literature.  In more general cases the equations are notably simpler than those previously encountered and they are conjectured to hold beyond the weak-field regime. Finally we derive the equations that determine the steady-states as well as the critical values of the control parameters beyond which a steady-state becomes unstable.

\end{abstract}
\begin{flushright}
\today
\end{flushright}
\end{titlepage}
\section{Introduction}

In 1999 the {\it Thermodynamic Field Theory} (TFT) was proposed, which describes the behavior of thermodynamic systems beyond the linear regime \cite{sonnino}-\cite{sonnino4}. This approach starts from the concept of {\it entropy production}, which can always be expressed  as a bilinear form of {\it some} variables characterizing the departure from equilibrium. If these variables are the thermodynamic forces, it is possible to formulate a covariant TFT and to find the {\it thermodynamic field equations} whose solutions give the generalized relation between the thermodynamic forces and the conjugate flows. 

In refs \cite{sonnino}-\cite{sonnino4}, the thermodynamic field equations have been obtained from three postulates:
\begin{description}
\item[1)] {\it The shortest path principle}
\item[2)] {\it The closedness of the thermodynamic field strength}
\item[3)] {\it The principle of least action}
\end{description}
\noindent where the thermodynamic field strength is determined from the skew-symmetric part of the tensor relating the thermodynamic forces with the conjugate flows. The second postulate reflects the observation that, in all cases examined so far, the sources of the thermodynamic fields {\it{i.e.}}, the internal fluctuations, time-dependent boundary conditions and external perturbations, only affect the balance equation of the dual form of $f$. This experimental observation allows us to identify the system of thermodynamic field equations as an analogue of Maxwell's equations. In the present note we will see that the second postulate is analogue to the absence of magnetic sources in a higher-dimensional version of Maxwell's equations. 

The validity of this theory in the weak-field approximation has been tested in many examples of $\alpha-\alpha$ or $\beta-\beta$ processes such as, for example, the thermoelectric effect and the unimolecular triangular chemical reaction (see refs \cite{sonnino} and \cite{sonnino1}). In the first case, the TFT provides new predictions when the material is subjected to strong electric field and, in the second, the thermodynamic field equations reproduce the well-known De Donder law. 

Transport processes in magnetically confined plasmas, which are of $\alpha-\alpha$ type, have also been analyzed using the TFT. In particular, in ref. \cite{sonnino5} the thermodynamic field equations were solved in the weak-field approximation of the classical and the Pfirsch-Schl{\"u}ter regimes. We found that the TFT does not correct the expressions for the ionic heat fluxes predicted by the neoclassical theory. On the other hand, the fluxes of matter and electronic energy (heat flow) are enhanced in the nonlinear classical and Pfirsch-Schl{\"{u}}ter regimes. This phenomenon would have been amplified had we used the strong (exact) thermodynamic field equations. These results are in line with the experimental observations.

The thermodynamic field equations, in the weak-field approximation, have also been applied to several $\alpha-\beta$ processes. For example the Field-K{\"{o}}r{\"{o}}s-Noyes model, in which the thermodynamic forces and flows are related by an asymmetric tensor, was analyzed in ref \cite{peeters}. Even in this case the numerical solutions of the model are in agreement with the theoretical predictions of the TFT. Recently, the Hall effect \cite{sonnino6}, \cite{hall} and, more generally, the galvanomagnetic and thermomagnetic effects have been analyzed in nonlinear regimes. In each of these papers it was shown that the TFT successfully describes the known physics in the nonlinear regimes and also predicts new interesting effects such as the nonlinear Hall effect. The theoretical predictions of the nonlinear Hall effect have been confirmed experimentally. In ref \cite{hall} it was shown that, for materials with low thermoelectric power coefficients and in the temperature range of available experimental data, the theoretical predictions of TFT agree with experiments. 

When the thermodynamic system approaches equilibrium, the solution of the field equations reduces to the Onsager-Casimir tensor and satisfies the Prigogine theorem of {\it minimum entropy production} \cite{prigogine1}. Far from equilibrium Glansdorff and Prigogine have demonstrated that, for time independent boundary conditions, a thermodynamic system relaxes to a steady-state satisfying the {\it Universal Criterion of Evolution} \cite{prigogine}. Glansdorff and Prigogine obtained this result in 1954 using a purely thermodynamical approach. As indicated in refs \cite{sonnino}-\cite{sonnino4}, making use of the second principle of thermodynamics, the Shortest Path Principle should ensure the validity of the Universal Criterion of Evolution.

In the present manuscript, using the coordinate independent language of Riemannian geometry, we reformulate the TFT in a generally covariant way. This allows us to extend several weak-field results to the strong-field regime.  After expressing the concepts of thermodynamic forces, conjugate flows and entropy production in this new language in sec \ref{entropy}, we go on to study the relaxation of a thermodynamic system in section \ref{riposo}.  In particular we prove that if the shortest path principle is valid, then the Universal Criterion of Evolution, written in a covariant form, is automatically satisfied. We also demonstrate that the term expressing  the Universal Criterion of Evolution, written in a covariant form, has a {\it local minimum} at the geodesic line. Physically this means that a thermodynamic system evolves towards a steady-state with the least possible dissipation. We shall refer to this proposal, together with its corollary the shortest path principle, as the {\it Minimum Dissipation Principle}. 

In section \ref{eqs} we obtain the thermodynamic field equations. For $\alpha-\alpha$ or $\beta-\beta$ processes we obtain the same equations found in refs \cite{sonnino}-\cite{sonnino4}, but for general processes, we propose a new set of the field equations, which are conjectured to be also valid beyond the weak-field regime. These equations are notably simpler than the ones found in ref.\cite{sonnino3}. In section \ref{stability}, we establish the equations that determine the steady-states and the critical values of the control parameters of a generic thermodynamic system at which a steady state becomes marginally stable. These equations are also conjectured to hold in the strong-field regime and therefore generalize the weak-field equations reported in refs \cite{sonnino2}-\cite{sonnino3}. Examples of calculations of the geometric quantities used in the new formulation of TFT can be found in the appendices. 

\section{Forces, Flows and Entropy Production}\label{entropy}

The central object of study in Thermodynamic Field Theory (TFT) is the thermodynamic space, which is a smooth, path-connected, geodesically complete, real manifold $M$ equipped with a Riemannian metric $g$.  A Riemannian metric is a positive-definite quadratic form on a manifold's tangent spaces $T_xM$, which varies smoothly from a point $x$ of the manifold $M$ to another.  The metric $g$ yields a smoothly-varying inner product
\beq
\langle A,B\rangle=A^\mu g_{\mu\nu} B^{\nu},\qquad A,B\in T_x M \label{iprod}
\eeq
where following the Einstein summation convention repeated indices are summed, as they will be in the remainder of this note.  The inner product allows one to define the lengths of curves, angles and volumes. 

$M$ contains one special point named $x=0$, which corresponds to the unique {\it equilibrium state}.  Note however that in practice the thermodynamical forces, which provide local coordinates on patches of M, will be not be independent but rather will satisfy a set of constraints that determines a submanifold $N$ of $M$.  When the submanifold $N$ of physical configurations does not include equilibrium the system relaxes do a different steady-state.

Given two points $x\in M$ and $y\in M$, we may construct a shortest path $\gamma (t)$ such that
\begin{equation}\label{vt1}
\gamma:[t_i,t_f]\rightarrow M\hsp \gamma(t_i)=x\hsp\gamma(t_f)=y
\end{equation}
The path $\gamma$ is automatically a geodesic.  Clearly for some manifolds and choices of $x$ and $y$ there will be multiple paths $\gamma$ that minimize the length, thus it is crucial that observable quantities be independent of the choice of path. The metric $g_{\mu\nu}$ allows to define the invariant {\it affine parameter} $\tau$: 
\begin{equation}\label{vt1a}
d\tau^2\equiv g_{\mu\nu}dx^{\mu}dx^{\nu}
\end{equation}
\noindent and our manifold $M$ is "metrized" defining the length of an arc $L$ by the formula
\begin{equation}\label{vt1b}
L=\int_{\tau_1}^{\tau_2}(g_{\mu\nu}\dot{x^{\mu}}\dot{x^{\nu}})^{1/2}d\tau
\end{equation}
\noindent The positive definiteness of the matrix $g_{\mu\nu}$ ensures that $d\tau^2\geq 0$ and allows us to choose an affine parameter $\tau$ that increases monotonically as the thermodynamic system evolves to a stable state. 

We introduce now the {\it exponential map} which is a map from the tangent space $T_xM$ of a Riemannian manifold $M$ at the point $x$ to another point $\textup{exp}_x(V)$ in $M$ (see for example ref. \cite{katok}). The point $\textup{exp}_x(V)$ is the point $\gamma_{V}(t_f)$ on the geodesic $\gamma$ that starts at ${\gamma}_{V}(t_i)=x$, where it is tangent to $\dot{{\gamma}}_{V}(t_i)=V$. $t_f$ is determined by the condition that the arc length along $\gamma$ from $x$ to $\textup{exp}_x(V)=\gamma_{V}(t_f)$ is the norm $\langle V,V\rangle^{1/2}$ of the vector $V$. This map is usually denoted by
\begin{equation}\label{vt1c}
\textup{exp}_x(V)=\gamma_{V}(t_f)
\end{equation}

We next introduce the generalized thermodynamic force vector $U_x\in T_xM$ at each point $x\in M$, which is defined along a shortest path $\bar\gamma_{U_x}$ by the relation
\begin{equation}\label{vt2}
\textup{exp}_x(U)=\bar\gamma_{U_x}(t_f)=0
\end{equation}
Therefore, the exponential map Eq.~(\ref{vt2}) associates to every tangent vector ${U}_x\in T_xM$ the equilibrium point $\gt_{U_x}(t_f)=0$ in $M$.  A thermodynamic system at $x$ reaches the equilibrium state after traveling along the shortest path $\bar\gamma_{U_x}$ with velocity proportional to $U_x$ at the point $x$. The norm $\mid U_x\mid$ indicates the geodesic distance from the point $x$ to the equilibrium point measured along the shortest path ${\bar\gamma}_{U_x}$. {\it The components $U_x^{\mu}$ of $U_x$ are interpreted as the generalized thermodynamic forces}. The exponential map is a one-to-one correspondence between a ball in $T_xM$ to a neighborhood of $x\in M$ and so Eq.~(\ref{vt2}) defines a {\it correspondence between the generalized thermodynamic forces and the points of manifold} $M$. From now on we shall omit the suffix $x$ being implicitly understood the dependence of vector $U$ on point $x$. In \ref{flat} and \ref{sphere} we will provide two examples of calculations of the map $\textup{exp}_x(U)$.  In the first example the thermodynamic space $M$ is flat space and so matrix tensor coincides with the Onsager matrix, while in the second $M$ is the two-sphere.  

The {\it thermodynamic flows} $J_{\mu}$, which are conjugate to the thermodynamic forces $U^{\mu}$, are defined by
\begin{equation}\label{vt3}
J_{\mu}=\lambda_{\mu\nu}U^{\nu}
\end{equation}
\noindent where $\lambda_{\mu\nu}$ is an asymmetric tensor and, for brevity, we have suppressed the dependence of $J$, $g$ and $U$ on the point $x$ and geodesic $\bar\gamma$. Any 2-tensor may be decomposed into a symmetric and antisymmetric piece. The symmetric piece of $\lambda$ is identified with the metric tensor $g$ and we shall name the skew-symmetric piece $f_{\mu\nu}$. Eq.~(\ref{vt3}) can then be re-written as
\begin{equation}\label{vt3a}
J_{\mu}=\lambda_{\mu\nu}U^{\nu}=(g_{\mu\nu}+f_{\mu\nu})U^{\nu}
\end{equation}
\noindent $J_{\mu}$ is a one form $J\in T^*M$ which, like $U$, only vanishes at the equilibrium point $x=0$ although there may be points at which both $U$ and $J$ depend on a choice of $\bar\gamma$. Eq.~(\ref{vt3a}) allows one to generalize the inner product $\langle A,B \rangle$ of two vectors $A$ and $B$ in $T_xM$ from Eq.~(\ref{iprod}) to
\begin{equation}
\langle A,B \rangle\equiv (g_{\mu\nu}+f_{\mu\nu})A^{\mu}B^{\nu}
\end{equation}

Finally we will define the entropy production $\sigma$ at the point $x$ to be
\begin{equation}\label{vt4}
\sigma=\langle U,U \rangle=(g_{\mu\nu}+f_{\mu\nu})U^{\mu}U^{\nu}=g_{\mu\nu}U^{\mu}U^{\nu}\geq 0
\end{equation}
\noindent where the inequality corresponds to the second law of thermodynamics. As the metric tensor $g_{\mu\nu}$ is a positive definite matrix, this inequality is always satisfied. The entropy production $\sigma$ is an observable quantity and so the consistency of the formulation demands that it be independent of the arbitrary choice of $\bar\gamma$.  In fact $\sigma$ is just the square of the length of the shortest geodesic from $x$ to $0$, and so it is independent of the choice among geodesics of equal length and thus is globally-defined as was required.  

In this note we will restrict our attention to homogeneous systems, therefore points $x\in M$ will correspond to configurations of the entire system.  In particular the entropy flux through the boundaries, will be constant.  This implies that any variation in the entropy production $\sigma$ is equal to the variation of the time derivative of the total entropy.  The generalization to inhomogeneous systems is straightforward and follows the strategy employed in Ref.~\cite{sonnino4}.

\section{Relaxation to a Steady-State}\label{riposo}

In the TFT description of a thermodynamic system a homogeneous configuration corresponds to a point $x$ in the thermodynamic space $M$.  The corresponding thermodynamic forces, which are completely determined by the configuration $x$, are assembled into the vector $U$ defined in Eq.~(\ref{vt2}).  If the point $x$ is not a steady-state (see section~\ref{stability}) then the system will evolve.  In this section we will consider the process known as {\it relaxation} in which the initial velocity vanishes and the system relaxes to a steady-state $y$.

The TFT description of relaxation rests upon two conjectures
\begin{description}
\item[a)] {\it The manifold ($M,g$) is a Riemannian manifold with metric $g$};
\item[b)] {\it The Minimum Dissipation Principle}.
\end{description}
The Minimum Dissipation Principle will be formulated in Subsec.~(\ref{MDP}), where we will see that it implies that during the process of relaxation the configuration traces out a geodesic $\gamma$ in thermodynamic space.  In Refs.~\cite{sonnino}-\cite{sonnino3} the geodesic property was referred to as the {\textit{shortest path principle}}. It allows one to write
\beq
x=\gamma(0),\qquad y=\gamma(t_f)=exp_x(V)
\eeq
where $V$ is a tangent vector to $\gamma$ at $x$ whose norm is the geodesic distance from $x$ to $y$.  The system begins at $x$ at time $0$ and reaches the steady-state $y$ at the time $t_f$.  The process of relaxation begins as soon as the system is released, that is, immediately after time zero.  In support of this conjecture, we will now argue that $V$ automatically satisfies a manifestly covariant form of the Minimum Entropy Production, the Glansdorff-Prigogine Universal Criterion of Evolution and the Minimum Dissipation Principle.

\subsection{The Minimum Entropy Production Theorem}\label{MEPT}

In 1945-1947, I. Prigogine proved the {\it Minimum Entropy Production Theorem} \cite{prigogine1}, which concerns the relaxation of thermodynamic systems near equilibrium.  This theorem states that:

\noindent {\textbf{Minimum Entropy Production Theorem:} } {\textit{Regardless of the type of processes considered, a thermodynamic system, near equilibrium, relaxes towards a steady-state $y$ in such a way that the inequality}
\begin{equation}\label{mept1}
\frac{d}{d t}(V_{\mu}V^{\mu})=\frac{d}{d t}\langle V,V\rangle\leq 0
\end{equation}
\noindent {\it is satisfied during throughout the evolution and is only saturated at $y$.} The Minimum Entropy Production Theorem has been proven by I. Prigogine, using a purely thermodynamical approach. We shall now demonstrate that this theorem is automatic in the framework of TFT. 

Consider a geodesic $\gamma(\tau)$ parametrized by $\tau\in [\tau_i,\tau_f]$ equipped with a tangent vector field $W$ normalized to have unit length
\beq\label{mept3}
W=\frac{\partial}{\partial \tau} \gamma(\tau) = \gamma^*(\frac{\partial}{\partial \tau}),\qquad \langle W,W\rangle=1
\eeq
Notice that $\tau$ is not necessarily the time $t$, as the velocity is not necessarily unity throughout the evolution.  The vector field $W$ evaluated at any point $x$ on the curve $\gamma$ is a vector $W(x)\in T_x\gamma\subset T_xM$.  We consider parameterizations of $\gamma$ such that $W(x)\neq 0$. We then have the following identity
\begin{equation}
\mathcal{L}_{W(\gamma(\tau))} \langle V,V\rangle=\frac{d}{d \tau}(V_{\mu}V^{\mu})
\end{equation}
\noindent where $\mathcal{L}_{W(\gamma(t))}$ denotes the Lie derivative with respect to the vector field $W$ and, since we are near equilibrium, indices are raised and lowered using the Onsager matrix which is a metric for Euclidean space. 

There always exists a neighborhood $U\subset M$ of $y$, which physically corresponds to the near-equilibrium approximation when $y$ is near equilibrium, such that we may uniquely identify the coordinates $x^\mu$ of a point $x$ with the vectors $V(x)$ in a subset of a tangent fiber.  One example of such a $U$ is a ball whose radius is the injectivity radius, which in topologically trivial situations is roughly the inverse square root of the curvature, but any $U$ such that each $x$ and $y$ in $U$ are connected by a unique geodesic will suffice.  To show that there is a correspondence between points and vectors we need to show that given $x$ we can determine $V(x)$ and vice versa.  We have already seen that we may uniquely define $V(x)$ from the unique geodesic in $U$ connecting $x$ and $y$.  

To choose a name for the coordinate $x$ given a choice of $V(x)$ we must first confront the fact that $V(x)$ is an element of the tangent space at $x$, $T_xM$, and so to compare different $V(x)$'s we must somehow place them in the same space.  This may be done canonically in $U$ using the fact that in $U$ there is a unique geodesic $\gamma_x$ from $x$ to the steady-state $y$.  Thus we may define the vector $A_y(x)\in T_yM$ by parallel transporting $V(x)$ from $x$ to $y$ along $\gamma$.  We may now use any choice of coordinates for $y$ to write $x$ on the same coordinate chart
\beq\label{equil1}
y-x=A_y(x)=V(x)
\eeq
where we have used parallel transport to define a coordinate system for $T_xM$ which allows us to identify $V(x)$ and $A$.  Notice that each $V(x)$ arises from a unique $x$ because $V(x)$ is parallel to $\gamma$ and the geodesic equation implies that parallel vectors remain parallel under parallel transport, and so $A$ is also parallel to $\gamma$.  Thus the uniqueness of the geodesic implies the uniqueness of $A$ and so by Eq.~(\ref{equil1}) the uniqueness of $x$.  The sign in Eq.~(\ref{equil1}) is a result of the fact that $V(x)$ points from $x$ to the point $y$. In these coordinates the metric is the identity at $y$, the Christoffel symbols vanish at $y$, the geodesic curves in $U$ that pass through $y$ are straight lines in $U$ and the entropy production is given by
\beq\label{equil2}
\sigma=|x|^2+O(x^3)
\eeq
\noindent Let us now evaluate the Lie derivative of $\langle V, V\rangle$ when the thermodynamic system relaxes from the point $x$ to the stationary point $y=\gamma (t_f)$.  According to TFT this relaxation follows the unique geodesic in $U$ that connects them, which is the straight line
\beq\label{equil6}
 x=y+V(\tau),\qquad V(\tau)=V(x)f(\tau)
\eeq
in the coordinates (\ref{equil1}). We then find
\beq\label{equil5}
\mathcal{L}_{W(\gamma(\tau))} \langle V(\tau), V(\tau)\rangle =
2\langle V(\tau),\frac{\partial}{\partial \tau}V(\tau)\rangle
=2f(\tau)f\p(\tau)\langle V(x),V(x)\rangle\leq 0
\eeq
\noindent where the last inequality is a consequence of the fact that $\langle V(x),V(x)\rangle$ is positive by the positive definiteness of the Onsager tensor $g$, while $f(\tau)$ is positive and $f\p(\tau)$ is negative because $f(\tau)$ is the length of $V$ which monotonically falls to zero during the relaxation.  Only at the steady-state $y$ is $f(\tau_f)=0$ and so the inequality is saturated.

\subsection{The Universal Criterion of Evolution}\label{UCE}

The minimum entropy production theorem is generally not satisfied far from equilibrium.  However in 1964 P. Glansdorff and I. Prigogine demonstrated that a similar theorem continues to hold for any relaxation to a steady-state. We report the original version of the Glansdorff and Prigogine theorem \cite{prigogine}: {\it When the thermodynamic forces and conjugate flows are related by a generic asymmetric tensor, regardless of the type of processes, for time independent boundary conditions, a thermodynamic system, even in strong non-equilibrium conditions, relaxes to a steady-state in such a way that the following Universal Criterion of Evolution is satisfied}
\begin{equation}\label{uce1}
(g_{\mu\nu}+f_{\mu\nu})V^{\nu}\frac{dV^{\mu}}{dt}=\langle V,\frac{dV}{dt}\rangle\leq 0\qquad\quad \Bigl(\langle V,\frac{dV}{dt}\rangle=0\quad\mathrm{at\ the\ steady\ state}\Bigr)
\end{equation}
\noindent Again Glansdorff and Prigogine have demonstrated this theorem using a purely thermodynamical approach. In this subsection we shall prove that the Shortest Path Principle implies that a covariant form of the Universal Criterion of Evolution is automatically satisfied for relaxation processes in TFT. Consider a geodesic $\gamma(t)$ parametrized by $\t\in [\t_i,\t_f]$ equipped with a tangent vector field $W$
\beq\label{uce2}
W=\frac{\partial}{\partial \t} \gamma(\t) = \gamma^*(\frac{\partial}{\partial \t}),\qquad \langle W,W\rangle=1\eeq
\noindent 
Again we have parametrized $\gamma$ so that the velocity vector $W$ always has a unit norm.  As the velocity of the relaxation is an observable and is not necessarily equal to one, we must stress that $\t$ refers to a particular parametrization and not necessarily to time.  However it is monotonically increasing with respect to time.  As usual we also define a second vector field $V$ on the curve $\gamma$ such that at any point $x$ the vector $V(x)$ satisfies
\beq\label{v}
\gamma(\t_f)=\textup{exp}_x(V(x))
\eeq
\noindent We will consider curves $\gamma$ which contain no closed loops, that is curves for which $\gamma(\t)=\gamma(\t\p)$ implies $\t=\t\p$ and so the condition (\ref{v}) uniquely defines the vector field $V$. 

Now, our task is to evaluate the term in Eq.~(\ref{uce1}) when the thermodynamic system relaxes towards a steady-state along a geodesic.  This term contains a time derivative of the vector $V$, which {\textit{a priori}} is ill-defined as $V$ at different times inhabits the tangent fibers over different points of $M$, which are not canonically isomorphic.  The time derivative may be defined using the Lie derivative with respect to the vector $W$.  This leads to the Covariant Form of the Universal Criterion of Evolution
\beq \label{uce1c}
\langle V,\mathcal{L}_W V\rangle\leq 0\qquad (\textup{with }\langle V,\mathcal{L}_W V\rangle=0 \textup{\ only at the steady state})
\eeq
From now on, the term $\langle V,\mathcal{L}_W V\rangle$ will be referred to as the {\it Glansdorff-Prigogine quantity}.  Notice that $V(\gamma(\t_f))=0$ and so in particular the Glansdorff-Prigogine quantity relaxes to zero
\beq
\langle V(\gamma(\t_f)),\mathcal{L}_{W(\gamma(\t_f))} V(\gamma(\t_f)\rangle=0.
\eeq

The quantity defined by the Lie derivative differs from the quantity in the minimum entropy production theorem, in fact the difference between these two terms is similar to the difference between the minimum entropy production term and the noncovariant time derivative term.  The decomposition for the noncovariant derivative is
\begin{eqnarray}\label{uce3}
\frac{d }{d \t} \langle V,V\rangle &=&(g_{\mu\nu}+f_{\mu\nu})V^{\nu}\frac{dV^{\mu}}{d\t}+V^{\mu}\frac{d}{d\t}
[(g_{\mu\nu}+f_{\mu\nu})V^{\nu}]
\nonumber\\
&=&\langle V,\frac{dV}{d\t}\rangle+V^{\mu}\frac{d}{d\t}[(g_{\mu\nu}+f_{\mu\nu})V^{\nu}]
\end{eqnarray}
and for the Lie derivative it is 
\begin{equation}\label{uce4}
\frac{d }{d \t}\langle V, V\rangle=\mathcal{L}_W\langle V, V\rangle=\langle V,\mathcal{L}_W V\rangle+V^{\mu}(\mathcal{L}_W {\bf (g+f)} V)_{\mu}
\end{equation}

\noindent We shall now prove the covariant form of the Universal Criterion of Evolution.

\noindent {\textbf{Theorem:} } {\textit{For all $\t<\t_f$ we have the strict inequality $\langle V,\mathcal{L}_{W} V\rangle <0$.}}

\noindent
{\textbf{Proof:}} For every point $x$ on the curve $\gamma$ the vectors $V(x)$ and $W(x)$ are elements of the one-dimensional vector space $T_x\gamma$, therefore they are parallel and there exists a function $f(\gamma(\t))$ such that
\beq
V(\gamma(\t))=f(\gamma(t))W(\gamma(\t)). \label{fw}
\eeq
We may now evaluate the Lie derivative
\bea
\mathcal{L}_{W(\gamma(\t))} V(\gamma(\t))&=& [W(\gamma(\t)),V(\gamma(\t))] = [W(\gamma(\t)),f(\gamma(\t))W(\gamma(\t))]\nonumber\\
&=&[\gamma^*(\frac{\partial}{\partial \t}),f(\gamma(\t))\gamma^*(\frac{\partial}{\partial \t})]=\frac{\partial f(\gamma(\t))}{\partial \t} W(\gamma(\t)). \label{lv}
\eea
Combining Eqs.~(\ref{fw}) and (\ref{lv}) we may evaluate $\langle V,\mathcal{L}_{W} V\rangle$ as a function of $f$
\beq
\langle V,\mathcal{L}_{W} V\rangle=\langle fW,\frac{\partial f}{\partial\t}W \rangle = f\frac{\partial f}{\partial \t}\langle W,W\rangle. \label{prodotto}
\eeq
The fact that $g$ is Riemannian and so positive definite implies that $\langle W,W\rangle>0$.  We have seen that $V(\gamma(\t_f))=0$ and we have imposed that $W\neq 0$ therefore $f(\t_f)=0$.  In addition the fact that $W\neq 0$ implies that the distance from $\gamma(\t)$ to $\gamma(\t_f)$ has no critical points and so $f$ has no critical points.  Thus $f$ is strictly monotonic in $\t$ and at the maximal value $\t=\t_f$ it vanishes.  As a result $f$ and $\partial f/\partial \t$ have opposite signs
\beq
f\frac{\partial f}{\partial\t}<0
\eeq
and so the right hand side of Eq.~(\ref{prodotto}) is the product of a positive and a negative term.  Therefore $\langle V,\mathcal{L}_{W} V\rangle$ is negative as desired. 

At this point one may object that the Glansdorff-Prigogine quantity in Eq.~(\ref{uce1c}) is not actually the quantity that appears in (\ref{uce1}) because $t$ and $\t$ are not equal and so $W$ is not actually the velocity of the system.  This implies that the covariant form of the Universal Criterion of Evolution is not in general equivalent to the original form.  However after a little algebra one can see that they are equivalent whenever the deceleration $|a|$, velocity $v$ and affine length $l$ of the system satisfy the following inequality
\beq
|a|<\frac{v^2}{l-\tau} \label{vincolo}
\eeq
at every affine time $\t$.  The acceleration is unconstrained, and so this is only a constraint on the dissipation of the system as it approaches the steady-state and slows.  However near the steady-state the denominator of the right hand side of Eq.~(\ref{vincolo}) shrinks to zero and so even near the end of the relaxation the constraint on the rate of dissipation appears to be quite weak.  This constraint will be compared with experiment in a subsequent publication.

\subsection{The Minimum Dissipation Principle}\label{MDP}

In refs \cite{sonnino}-\cite{sonnino3} we have introduced the {\it Shortest Path Principle}. In Subsec~\ref{UCE}, we have demonstrated that if the shortest path principle is valid then the Universal Criterion of Evolution, written in a covariant form, is automatically satisfied. In this subsection we will show that the expression $\langle V,\mathcal{L}_W V\rangle$ has a {\it local minimum} at the geodesic line. We refer to this theorem, together with the Shortest Path Principle, as the {\it Minimum Dissipation Principle}. The Minimum Dissipation Principle corresponds to the following physical conjecture:
\vskip0.3cm
\noindent {\it For any type of physical process, even in strong non-equilibrium conditions, a thermodynamic system subjected to time-independent boundary conditions relaxes towards a steady-state following the path with the least possible dissipation}. 
\vskip0.3cm
Let us consider an arbitrary path $\gamma(t)$ parameterized by $t\in [0,t_f]$ equipped with a tangent vector field
\beq\label{mdp1e}
W=\frac{\partial}{\partial t} \gamma(t) = \gamma^*(\frac{\partial}{\partial t})
\eeq
This time we fix the velocity of the trajectory, this is necessary in any minimization of dissipation because otherwise one could always trivially speed up the dissipation by accelerating the process.  In particular we will parametrize $\gamma(t)$ by imposing that $W$ is a unit vector
\beq\label{mdp2e}
\langle W,W\rangle=1
\eeq
At every time $t$ there exists at least one geodesic $\alpha_t(\tau)$ such that
\beq\label{mdp3e}
\alpha_t(0)=\gamma(t)\hsp\alpha_t(\tau_f)=\gamma(t_f)
\eeq
There also exists a vector $V_t$ in the tangent space $T_{\gamma(t)}\alpha_t\subset T_{\gamma(t)}M$ such that
\beq\label{vt}
\gamma(t_f)=\textup{exp}_{\gamma(t)}(V_t)
\eeq
Eq.~(\ref{vt}) defines the vector $V_t$ at every point $\gamma(t)$ on the curve $\gamma$ and so it defines a vector field $V(\gamma(t))$ on $\gamma$ with vectors in the tangent space $TM$.  Thus we may define the Lie derivative $\mathcal{L}_{W(\gamma(t))} V(\gamma(t))$ which will be another vector field on $\gamma$.

\begin{figure*}[htb] 
\hspace{3cm}\includegraphics[width=10cm,height=12cm]{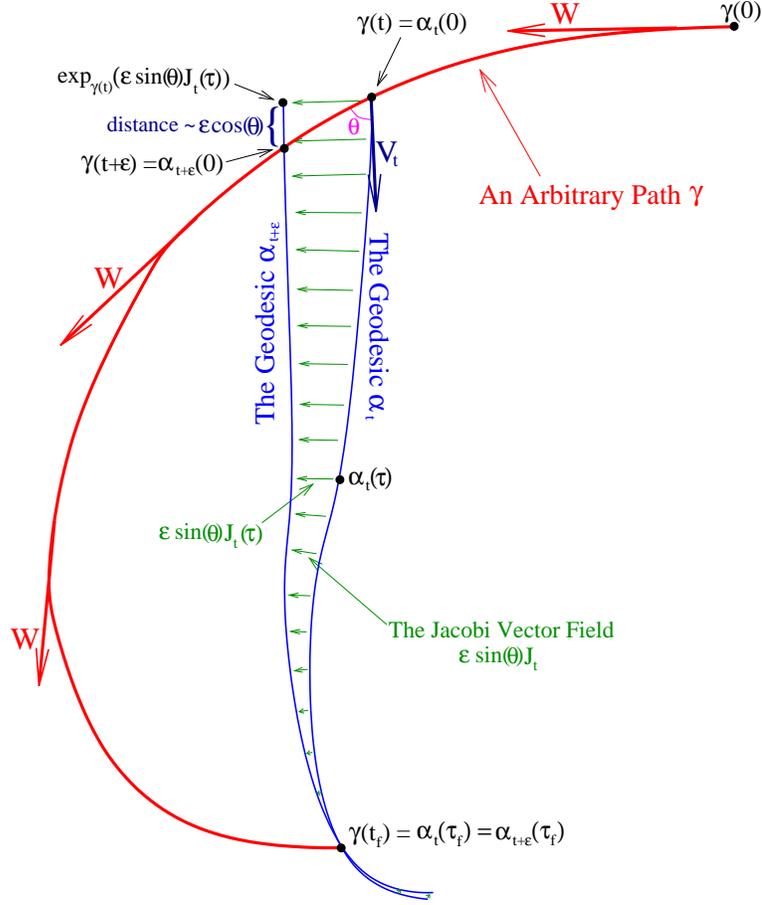}
\caption{\label{mdp}
As one moves along the curve $\gamma$ the minimal distance to $\gamma(t_f)$ shrinks at a speed of $cos(\theta)$.  This speed, which is proportional to the negative quantity $\langle V,\mathcal{L}_W V\rangle$, is locally minimized when $\theta=0$ corresponding to the case in which $\gamma$ follows a geodesic.}
\end{figure*}
  
We will show that, using the above parametrization
\beq\label{cos}
\langle V,\mathcal{L}_W V\rangle =-|V|\textup{cos}(\theta) 
\eeq
where $\theta$ is the angle between $\alpha$ and $\gamma$. This quantity is minimized when $\theta=0$, corresponding to the case in which $\alpha$ and $\gamma$ are parallel. Therefore the Glansdorff-Prigogine quantity is minimized everywhere iff $\alpha$ and $\gamma$ coincide everywhere, in which case $\gamma$ is a geodesic.

To prove this theorem we shall perform our calculations in Fermi coordinates \cite{fermi}. In particular, we shall work in the Fermi coordinate system such that the first derivatives of the components of the metric along the geodesic $\alpha_t$ vanish
\beq\label{mdp4}
\frac{\partial}{\partial x^{\lambda}} g_{\mu\nu}(\alpha_t)=0
\eeq
Therefore in Fermi coordinates, a distance $\epsilon$ from the geodesic, the components of the metric will differ by a correction of order $\epsilon^2$:
\begin{equation}\label{mdp5}
g_{\mu\nu}(\alpha_{t+\epsilon}) =g_{\mu\nu}(\alpha_{t})+\frac{\partial g_{\mu\nu}(\alpha_t)}{\partial x^{\lambda}}\epsilon^{\lambda}+\frac{1}{2}\frac{\partial^2 g_{\mu\nu}(\alpha_t)}{\partial x^{\lambda}\partial x^{\kappa}}\epsilon^{\lambda}\epsilon^{\kappa}+O(\epsilon^3)
\end{equation}
\noindent or 
\begin{equation}\label{mdp6}
\Delta g_{\mu\nu}= g_{\mu\nu}(\alpha_{t+\epsilon}) -g_{\mu\nu}(\alpha_{t})=\frac{\partial^2 g_{\mu\nu}(\alpha_t)}{\partial x^{\lambda}\partial x^{\kappa}}\epsilon^{\lambda}\epsilon^{\kappa}+O(\epsilon^3)
\end{equation}
\noindent In Fermi coordinates, we may interpret the Glansdorff-Prigogine quantity (\ref{cos}) using the fact that $\langle V,V\rangle$ is the length squared of the geodesic $\alpha_t$ and its Lie derivative with respect to $W$ is
\begin{equation}\label{mdp7}
\mathcal{L}_W\langle V,V\rangle=2\langle V,\mathcal{L}_W V\rangle + V^\mu V^\nu (\mathcal{L}_W {\bf g})_{\mu\nu}
\end{equation}
 \noindent Notice that the term $\mathcal{L}_W\langle V,V\rangle$ {\it does not} correspond to the Lie derivative of the entropy production, which as seen in the section \ref{entropy}, is defined to be $\mathcal{L}_W\langle U,U\rangle$. 

We shall now show that the second term on the right-hand side is of order $\epsilon$ while the other two terms are of order one
\beq
V^\mu V^\nu (\mathcal{L}_W {\bf g})_{\mu\nu}\sim O(\epsilon),\qquad \mathcal{L}_W\langle V,V\rangle\sim 2\langle V,\mathcal{L}_W V\rangle\sim O(1)
\eeq
Up to now, we have defined the vector $W$ only along the curve $\gamma$. However, Eq.~(\ref{mdp7}) requires a definition of the derivative of $W$ in directions that are not parallel to the curve.  Such derivatives have not yet been defined, but they may be defined if we extend the definition of $W$ to a neighborhood $U$ of the point $\alpha_t(0)$. We define $W$ at a point $\alpha_t(0)+\epsilon^{\lambda}$ in this neighborhood by parallel transporting $W(\alpha_t(0))$ along the unique geodesic $\delta x^{\lambda}=\epsilon^{\lambda}$  in $U$ that connects $\alpha_t(0)$ and $\alpha_t(0)+\epsilon^{\lambda}$ \cite{weinberg} (see Fig. (\ref{mdp2})):
 
\begin{figure}[ht]
\begin{center}
\leavevmode
\epsfxsize 10  cm
\epsffile{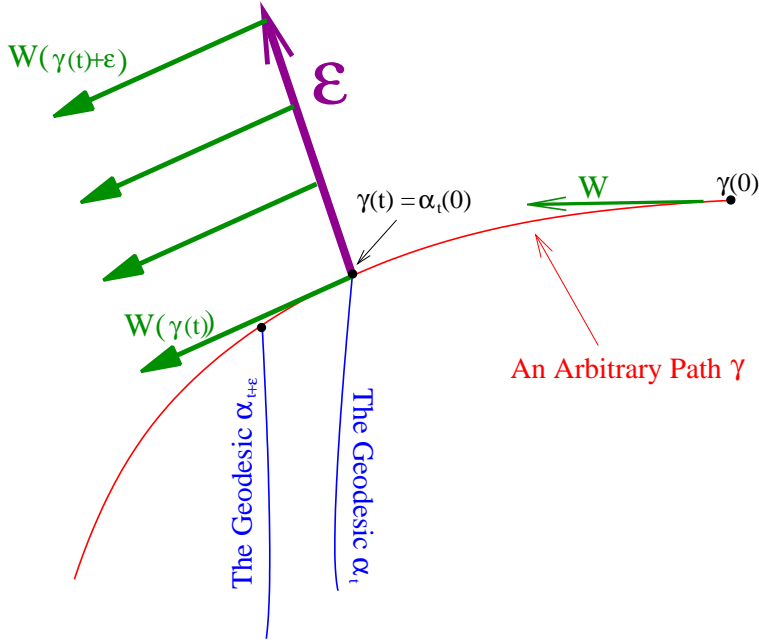}
\end{center}
\caption{\label{mdp2}
Definition define the vector $W$ in the neighborhood of the point $\gamma(t)$ by parallel transporting the vector $W(\gamma(t))$ along an edge $\delta x^{\lambda}=\epsilon^{\lambda}$.}
\end{figure}

\begin{equation}\label{mdp8}
W^{\mu}_{\epsilon}(x)=W_0^{\mu}+\Gamma_{\lambda\kappa}^{\mu}(x)\epsilon^{\lambda}W_0^{\kappa}+O(\epsilon^2)
\end{equation}

\noindent Eq.~(\ref{mdp8}) allows us to generalize the definition of the quantity $\langle V,\mathcal{L}_WV\rangle$ to arbitrary curves. The last term of Eq.~(\ref{mdp7}) is the derivative of the metric along the direction parallel to $\alpha_t$. It is easy to check that this term is of order $\epsilon$ in our coordinate system:
\begin{equation}\label{mdp9}
V^{\mu}V^{\nu}(\mathcal{L}_W{\bf g})_{\mu\nu}=V^{\mu}V^{\nu}W^{\lambda}\frac{\partial g_{\mu\nu}}{\partial x^{\lambda}}+2V_{\mu}V^{\nu}\frac{\partial W^{\mu}}{\partial x^{\nu}}
\end{equation}
\noindent or
\begin{equation}\label{mdp10}
V^{\mu}V^{\nu}(\mathcal{L}_W{\bf g})_{\mu\nu}=2V_{\mu}V^{\nu}W^{\lambda}\Gamma^{\mu}_{\nu\lambda}+2V_{\mu}V^{\nu}
\frac{\partial \Gamma_{\lambda\kappa}^{\mu}}{\partial x^{\nu}}\epsilon^{\lambda}W_0^{\kappa}+O(\epsilon^2)
\end{equation}
\noindent where we have used the identity
\begin{equation}\label{mdp11}
\frac{\partial g_{\mu\nu}}{\partial x^{\lambda}}=\Gamma^{\kappa}_{\lambda\mu}g_{\kappa\nu}+\Gamma^{\kappa}_{\lambda\nu}g_{\kappa\mu}
\end{equation}
\noindent However in Fermi coordinates, a distance $\epsilon$ from the curve the components of the affine connection will differ by a correction of order $\epsilon$:
\begin{equation}\label{mdp12}
\Gamma_{\mu\nu}^{\lambda}(\alpha_{t+\epsilon})=\Gamma_{\mu\nu}^{\lambda}(\alpha_{t})+\frac{\partial \Gamma_{\mu\nu}^{\lambda}(\alpha_{t})}{\partial x^{\kappa}}\epsilon^{\kappa}+O(\epsilon^2)=\frac{\partial \Gamma_{\mu\nu}^{\lambda}(\alpha_{t})}{\partial x^{\kappa}}\epsilon^{\kappa}+O(\epsilon^2)
\end{equation}
\noindent Observe that from Eq.~(\ref{mdp12}), we also have that 
\begin{equation}\label{mdp13}
\frac{\partial \Gamma_{\mu\nu}^{\lambda}(\alpha_{t})}{\partial x^{\eta}}=\lim_{\epsilon^\eta\rightarrow 0}\frac{\Gamma_{\mu\nu}^{\lambda}(\alpha_{t+\epsilon})-\Gamma_{\mu\nu}^{\lambda}(\alpha_{t})}{\epsilon^\eta}\sim O(1)
\end{equation}
Therefore, the second term on the right-hand side of Eq.~(\ref{mdp6}) is of order $\epsilon$. As a check, we note that Eq.~(\ref{mdp13}) reveals that also the Riemannian curvature tensor $R^{\mu}_{\nu\lambda\kappa}$ is of order $O(1)$, as it must be since our choice of coordinates cannot affect the curvature.

Now, we shall prove that the derivative of the length squared of $\alpha_t$ along the curve $\gamma$ is of order $O(1)$. For this, we have to compare the lengths of the geodesics $\alpha_t$ and $\alpha_{t+\epsilon}$. This leads us to the notion of Jacobi fields (see, for example, ref. \cite{frankel}). Let us consider a single infinite family of geodesics in the manifold $M$. Let $\tau$ be a parameter varying along each geodesic of the family, and let $t$ a parameter constant along each geodesic of the family, but varying as we pass from a geodesic to another. A Jacobi field is a vector field on a geodesic that describes infinitesimal deformations that interpolate between different geodesics in a one-parameter family. For example, the family $\alpha_t$ yields the Jacobi field $J_t$
\beq\label{mdp14}
J^{\mu}_t(\tau)=\frac{\partial\alpha^{\mu}_t(\tau)}{\partial t}
\eeq
on each geodesic $\alpha^{\mu}_t(\tau)$. 

Consider now two nearby geodesics in the $\a$ family, $\alpha_t$ and $\alpha_{t+\epsilon}$. We call the points $A$ on $\alpha_t$ and $A'$ on $\alpha_{t+\epsilon}$ corresponding points if they have equal values of $\tau$ (see Fig.~\ref{mdp3}). Let $\lambda_t^\mu=\epsilon\sin\theta J^\mu_t$ be the infinitesimal Jacobi vector (see Fig.~\ref{mdp}) such that \cite{synge}
\begin{equation}\label{mdp15}
g_{\mu\nu}\alpha_t^{\mu}\lambda_t^{\nu}=0
\end{equation}
This equation tells us that the deviation $\lambda_t^\mu$ is perpendicular to $\alpha_t^\mu$. The vector $\lambda_t^{\mu}$ satisfies the Jacobi equation \cite{synge}
\beq\label{jac}
\frac{D}{D\tau^2}\lambda_t^{\mu}(\tau)+R^{\mu}_{\nu\lambda\kappa}\frac{\partial{\alpha}_t^{\nu}(\tau)}
{\partial\tau}\lambda_t^{\lambda}(\tau)\frac{\partial{\alpha}_t^{\kappa}(\tau)}{\partial\tau}=0 
\eeq
where $D$ denotes the covariant derivative. Thus both terms in the Jacobi equation (\ref{jac}) are linear in $\epsilon$.  This implies that, in the limit $\epsilon\rightarrow 0$, the second derivative of our Jacobi field goes to zero. The relative angle $\phi$ at the point $A'$ (see Fig.~\ref{mdp3}), between $\alpha_t$ and $\alpha_{t+\epsilon}$, measured using parallel transport along the flow of the Jacobi field, can be easily evaluated taking into account the relation \cite{synge}
\begin{equation}\label{mdp17}
(\Delta\alpha_t^\mu)_{A'}(Y_\mu)_{A'}=\int\int Y_\mu R^{\mu}_{\nu\lambda\kappa}\alpha_t^\nu\frac{\partial\alpha_t^\lambda}{\partial\tau}\lambda_t^\kappa dt d\tau
\end{equation}
\noindent where $Y_{\mu}$ is a vector chosen arbitrarily at $A'$. Eq.~(\ref{mdp17}) shows then that $\phi$ is of order $\epsilon$. 

\begin{figure*}[htb] 
\hspace{1cm}\includegraphics[width=12cm,height=7cm]{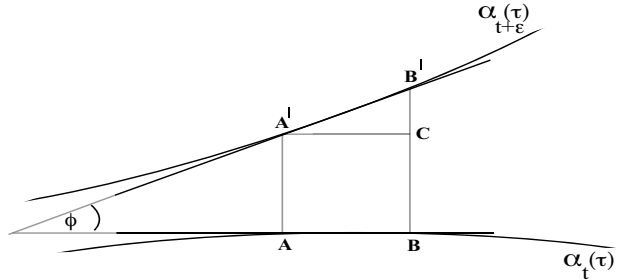}
\caption{\label{mdp3}
When the relative angle $\phi$ at the point $A'$ between $\alpha_t$ and $\alpha_{t+\epsilon}$ is of order $\epsilon$, the difference between length ${\bar {A'B'}}$ and length ${\bar {AB}}$ is of order $\epsilon^2$. Indeed, we have ${\bar {A'B'}}-{\bar {AB}}\simeq\frac{1}{2}\phi{\bar {CB'}}\sim O(\epsilon^2)$.}
\end{figure*}

Our task is to compute the limit of quantity 
\begin{equation}\label{mdp18}
\lim_{\epsilon\rightarrow 0}\frac{\Delta\langle V,V\rangle}{\epsilon}=2g_{\mu\nu}V^{\nu}\lim_{\epsilon\rightarrow 0}\frac{\Delta V^{\mu}}{\epsilon}+ V^{\mu}V^{\nu}\lim_{\epsilon\rightarrow 0}\frac{\Delta g_{\mu\nu}}{\epsilon}
\end{equation}
\noindent Using the relation (\ref{mdp6}), the last term on the right-hand side vanishes. As seen in Fig.~\ref{mdp3}, to leading order in $\epsilon$, the difference between the vector $V^{\mu}$ at the geodesic $\alpha_t$ and its image on the curve $\alpha_{t+\epsilon}$ under $exp_{\alpha_t}(\epsilon \textup{sin}(\theta) J_t)$ is of order $\epsilon^2$. In conclusion, the limit of the difference between the squared length of the geodesic $\alpha_{t+\epsilon}$ and the portion of the geodesic $\alpha_t$ such that $\alpha_{t+\epsilon}=\exp_{\alpha(t)}(\epsilon sin(\theta) J_t)$, vanishes.

\noindent However, the curve $\alpha_{t+\epsilon}$ is not quite the same as $exp_{\alpha_t}(\epsilon \textup{sin}(\theta) J_t)$, because we have parameterized $\alpha_{t+\epsilon}$ such that (see Fig.~\ref{mdp})
\beq\label{mdp19}
\alpha_{t+\epsilon}(\tau=0)=\gamma(t+\epsilon)\neq \exp_{\gamma(t)}(\epsilon sin(\theta) J_t(0)).
\eeq
That is to say that the Jacobi field is not parallel to $\gamma$, so it does not take the beginning of the curve $\alpha_t$ to the beginning of the curve $\alpha_{t+\epsilon}$. Instead $\exp$ of the Jacobi field $\epsilon \textup{sin}(\theta) J_t(0)$ takes the beginning of the curve $\alpha_t$ to a point $-\epsilon \textup{cos}(\theta)$ on the curve $\alpha_{t+\epsilon}$.  Up to corrections of order $\epsilon^2$ the curves $\alpha_t$ and $\alpha_{t+\epsilon}$ are the same length, and so the distance from $\alpha_{t+\epsilon}(0)$ to $\gamma(t_f)$ is that of $\alpha_t$ minus the correction $\epsilon \textup{cos}(\theta)$ that we have just calculated.  The distance squared then differs by $|V|\epsilon \textup{cos}(\theta)$.  By definition the Lie derivative is this quantity divided by $\epsilon$, which yields the claim (\ref{cos}).

\section{The Thermodynamic Field Equations} \label{eqs}

As seen in Eq.~(\ref{vt4}), the second principle of thermodynamics is satisfied if the metric $g$ is positive definite.  However the postulates that we have presented so far are insufficient to calculate $g$. 
In this section we will add a third postulate to TFT stating that the metric is the solution to a set of field equations which are in turn equal to the variations of an action functional.  In short we include the postulate:
\begin{description}

\item[c)] {\it The Principle of Least Action}.
\end{description}
\noindent In particular we will assert that the action is invariant with respect to general coordinate transformations and we will not include a cosmological constant term, as there has been no experimental evidence for the presence of such a term in TFT.  

We will also place the following three restrictions on our physical systems
\begin{description} 
\item[1)] {\it Internal fluctuations are neglected};
\item[2)] {\it Boundary conditions are time-independent};
\item[3)] {\it The system is not subjected to external perturbations}.
\end{description}
If these three conditions are simultaneously satisfied then the equations of motion will be {\it source-free}. The internal fluctuations can be analyzed using the Landau-Lifshitz theory and systems submitted to time-dependent boundary conditions have been examined by York in ref. \cite{york}. Examples of systems subjected to external perturbations, those most commonly encountered in practice, can be found in ref. \cite{lefever}. 

\subsection{$\alpha$-$\alpha$ and $\beta$-$\beta$ Processes}\label{TFEaa}

In $\alpha$-$\alpha$ and $\beta$-$\beta$ processes the matrix $\lambda_{\mu\nu}$ that relates the thermodynamic forces and flows (\ref{vt3}) is symmetric and so is equal to the metric $g$.  We therefore expect the action to be constructed entirely from $g$.  After the cosmological constant, the invariant constructed using the least number of derivatives of $g$ is the Ricci scalar $R\equiv g^{\mu\nu}R_{\mu\nu}$ where $R_{\mu\nu}$ is the Ricci tensor \cite{lovelock}
\begin{equation}\label{plasma9}
R_{\mu\nu}\equiv \Gamma_{\mu\lambda,\nu}^{\lambda}-\Gamma_{\mu\nu,\lambda}^{\lambda}+
\Gamma_{\mu\lambda}^{\eta}\Gamma_{\eta\nu}^{\lambda}-\Gamma_{\mu\nu}^{\eta}\Gamma_{\eta\lambda}^{\lambda}
\end{equation}
\noindent Of course one could add higher powers of $R$ and its derivatives to the action and maintain general covariance, but so far Eq.~(\ref{plasma9}) has been sufficient to reproduce the experimental data and numerical simulations. If such terms are indeed present in the action then, by dimensional analysis, they will be multiplied by characteristic scales of positive spatial dimension and they will be negligible at smaller distance scales.  Thus it may well be that above some threshold length scale, corresponding to very strong fields, higher order terms in $R$ will begin to play a role.  For example in gravitational theories such terms may play a role deep inside of black holes and immediately after the big bang, but are irrelevant at the scales of all observed phenomena.  By contrast in general relativity the nonlinearity of the $R$ term considered here already provides a significant correction to the clocks of the GPS satellites and its effect on Mercury's orbit was observed a hundred years ago.

We therefore postulate that the action $I$ takes the simple form
\begin{equation}
I=\int\sqrt{g}R_{\mu\nu}g^{\mu\nu}d\Omega\label{plasma10}
\end{equation}
\noindent where $d\Omega$ is a volume element and the domain of integration is the entire manifold $M$. Here $g$ denotes the determinant of the metric $g_{\mu\nu}$. The principle of least action implies that the variation of $I$ with respect to the metric vanishes, which yields the familiar thermodynamic field equations:
\begin{equation}\label{plasma11}
R_{\mu\nu}-\frac{1}{2}Rg_{\mu\nu}=0
\end{equation}
\noindent For $n\neq 2$ Einstein's equation (\ref{plasma11}) reduces to the Ricci flatness condition
\begin{equation}\label{plasma12}
R_{\mu\nu}=0
\end{equation}

\subsection{The General Case}\label{TFEab}

Let us now consider the general case in which $\lambda_{\mu\nu}$ is not necessarily symmetric 
\beq
J_{\mu}=\lambda_{\mu\nu}U^{\nu}=(g_{\mu\nu}+f_{\mu\nu})U^{\nu}\label{TFEab1}
\eeq
Any completely skew-symmetric tensor of type ($0,2$) defines a 2-form $\tilde f$. Given a skew-symmetric tensor $f$ of type ($2,0$) with components $f^{\mu\nu}$, we can define an ($n-2$)-form $\ast{\tilde f}$ called the Hodge dual of $f$. In terms of components, the relations between $\ast{\tilde f}$ and $f$ is
\begin{equation}\label{TFEab2}
\ast{\tilde f}_{j_1j_2\dots J_{n-2}}\equiv \frac{1}{2!}g^{1/2}\varepsilon_{\mu\nu j_1j_2\dots j_{n-2}}f^{\mu\nu}
\end{equation}
\noindent where $\varepsilon_{\mu\nu j_1j_2\dots j_{n-2}}$ is the Levi-Civita symbol:
\begin{equation}\label{TFEab3}
\varepsilon_{i,j\dots n}=\varepsilon^{i,j\dots n}=\left\{ \begin{array}{ll}
+1 & \mbox{if $ij\dots n$\ {\rm is\ an\ even\ permutation\ of}\ $1,2\dots ,n$}\\
-1 & \mbox{if $ij\dots n$\ {\rm is\ an\ odd\ permutation\ of}\ $1,2\dots ,n$}\\
0 &\mbox{{\rm otherwise}}
\end{array}
\right.
\end{equation}
\noindent The inverse of Eq. (\ref{TFEab2}) reads
\begin{equation}\label{TFEab4}
f^{\mu\nu}=\frac{1}{(n-2)!}g^{1/2}\varepsilon^{j_1j_2\dots j_{n-2}\mu\nu}(\ast{\tilde f})_{j_1j_2\cdots j_{n-2}}
\end{equation} 

To write the kinetic term for the $f$-field we will need to first define its antisymmetric 3-tensor field strength $H$
\begin{equation}\label{TFEab5}
H_{\mu\nu\rho}=\partial_{\mu}f_{\nu\rho}+\partial_\nu f_{\rho\mu}+\partial_\rho f_{\mu\nu}
\end{equation}
\noindent or simply 
\begin{equation}\label{TFEab5a}
\tilde H=d\tilde f
\end{equation}
\noindent where $d$ is the exterior derivative.  The square of the exterior derivative, $d^2$ contains the contraction of two ordinary derivatives with the $\epsilon$ tensor and so it vanishes by antisymmetry.  Therefore $H$ is closed
\begin{equation}\label{TFEab6}
d\tilde H=d^2\tilde f=0 
\end{equation}
\noindent 
In principle the balance equation for $\tilde H$ could also contain a source term $\tilde \mj$ 
\begin{equation}\label{TFEab8}
d\tilde H=\tilde \mj
\end{equation}
\noindent where $\mj$ is a closed 4-form, $d\tilde \mj=0$. 

Given a third-rank skew-symmetric tensor $H$ we can define an ($n-3$)-form $\ast{\tilde H}$, the Hodge dual of $H$, with components
\begin{equation}\label{TFEab9}
\ast{\tilde H}_{j_1j_2\dots J_{n-3}}\equiv \frac{1}{3!}g^{1/2}\varepsilon_{\mu\nu\lambda j_1j_2\dots j_{n-3}}H^{\mu\nu\lambda}
\end{equation}
As we shall see shortly, $\ast{\tilde H}$ also satisfies a balance equation with a source term, which we shall denote $\ast{\tilde J}$. This source term may be nonzero only if the three restrictions on our physical system are relaxed.  In other words internal fluctuations, external perturbations and time-dependent boundary conditions can serve as $\ast{\tilde J}$ sources for $\ast{\tilde H}$.  However in all cases examined so far no potential $\mj$ sources have been found for ${\tilde H}$. 

We are therefore led to the following postulate \cite{sonnino2},\cite{sonnino3}:
\begin{description}
\item[d)]{\it The 3-form $\tilde H$ has no sources} or, equivalently, {\it the 3-form $\tilde H$ is closed}.
\end{description}
\noindent Notice that, from the mathematical point of view, to impose $\tilde \mj=0$ it is sufficient to require that the tensor field $f$ does not possess any singularities and so Eq.~(\ref{TFEab5a}) is well-defined everywhere.  

We will now present an invariant action $I$ for the thermodynamic system.  As in the previous subsection, we will restrict our attention to terms that contain the least number of derivatives, as higher derivate terms will be suppressed by a characteristic length scale and so may only be relevant in the presence of very strong fields.  Again the metric may enter via an Einstein-Hilbert term equal to the Ricci scalar $R$.  The antisymmetric tensor $f$ may enter via a 2-derivative term $\tilde{H}\wedge *\tilde{H}$ generalizing the Maxwell action and also via a 0-derivative mass term $\chi\tilde{f}\wedge*\tilde{f}$ where $\chi$ is a dimensionful parameter.  Here $\wedge$ is the wedge product of differential forms and $*$ is the Hodge star defined in Eqs.~(\ref{TFEab2}) and (\ref{TFEab9}).  A cosmological constant term and more complicated functions of the mass term are not necessary to explain the data and so will not be included. In components the proposed Lagrangian density is then
\beq\label{TFEab11}
\mathcal{L}_{kin}=\frac{1}{12}H_{\mu\nu\rho}H^{\mu\nu\rho}+\frac{\chi}{2} f_{\mu\nu}f^{\mu\nu}
\eeq
where indices are raised using $g$.  The mass parameter $\chi$ is an {\it inverse of the characteristic length squared} of the thermodynamic system. 

The source $J_{\mu\nu}$ can be included to the theory by adding an interaction term to the Lagrangian density
\beq\label{TFEab12}
\mathcal{L}_{int}=S_{n-1}J_{\mu\nu}f^{\mu\nu}
\eeq
where $S_{n-1}=2\pi^{n/2}\Gamma (n/2)$ is the area of a unit $(n-1)$-sphere in $n$ space dimensions ($\Gamma (n/2)$ is the Gamma function).  Notice that if we do not add the source $J$, we could add a new term coupling the source $\mj$ to the dual field $f^{(dual)}$ where $df^{(dual)}=*H$. However, $f$ and $f^{(dual)}$ are not mutually local, since $f^{(dual)}$ at a point cannot be constructed from $f$ at that point together with a finite number of derivatives and so no local theory can simultaneously contain both $\mj$ and $J$.  In line with our postulate, we impose $\mj=0$ and so we only consider the source term in Eq.~(\ref{TFEab12}).

Combining (\ref{TFEab11}), (\ref{TFEab12}) and the Einstein-Hilbert action we arrive at the total action 
\beq\label{TFEab13}
I=\int\sqrt{g}\mathcal{L}=\int(\sqrt{g}R+\frac{1}{12}\tilde H\wedge\star{\tilde H}+\frac{\chi}{2} \tilde f\wedge\star{\tilde f}+S_{n-1}\tilde J\wedge\star\tilde{f})
\eeq

The relative normalization of the first two terms in the action (\ref{TFEab13}) may be modified by rescaling the metric, here we have chosen the convention known as the string frame in string theory \cite{string}. 

$*J$ will be treated as an external source, and so there are two variables whose variations yield the field equations.  The variation with respect to the metric yields a modified Einstein's equation
\beq\label{TFEab14}
R_{\mu\nu}\!-\!\frac{1}{2}g_{\mu\nu}R\!=\!-\frac{1}{4}(H_{\mu\lambda\rho}H_\nu^{\lambda\rho}\!-\!\frac{1}{6}g_{\mu\nu}H_{\lambda\kappa\rho}H^{\lambda\kappa\rho})\!-\!\chi\!\Bigl(f_{\mu\kappa}f_{\nu}^{\ \kappa}\!-\!\frac{1}{4}g_{\mu\nu}f_{\lambda\kappa}f^{\lambda\kappa}\Bigr)\!+\!\frac{S_{n-1}}{2}\!g_{\mu\nu}J_{\lambda\kappa}f^{\lambda\kappa}
\eeq
while the variation with respect to the $f$-field yields
\beq\label{TFEab15}
d*\tilde H+\chi *\tilde f=-S_{n-1}*\tilde J 
\eeq

We have seen that the antisymmetric tensor $f$ modifies the field equations for the metric.  One may wonder if it also affects the shortest path principle.  In general a $p$-tensor potential can be covariantly coupled to a $p$-dimensional $J$-source and to a $(d-p-2)$-dimensional $\mj$-source.  For example in 4-dimensional Maxwell theory the 1-tensor potential couples to electric and magnetic particles, while in 5 dimensions the electric sources are still particles but the magnetic monopoles are strings. In our case $p=2$, so the $J$-source is a string and the $\mj$-source is $(d-4)$-dimensional. In particular, the trajectories in the manifold of thermodynamic configurations are, being paths, 1-dimensional and so they cannot contain the $J$-source.  Except in 5-dimensions they also cannot contain the $\mj$-source. Thus, there is no covariant coupling of the trajectories to the $f$-field, as there would have been for an ordinary vector potential. 

Concretely, this means that the action for a relaxing configuration is proportional to the proper length of its trajectory
\beq\label{TFEab16}
S\propto\int d\tau \Bigl(\frac{dX^{\mu}}{d\tau}\frac{dX_{\mu}}{d\tau}\Bigl)^{1/2}
\eeq
just as in the case with no $f$-field.  The proper length of a path, by definition, is extremized by a geodesic.  Thus the solutions of the classical equations of motion for the evolution of a state are still geodesics in thermodynamic space, even when we include the $f$ field.  This is not to say that the solutions with an $f$ field are just the solutions to the system without the $f$-field, because the new Einstein equation (\ref{TFEab14}) no longer admits Ricci-flat metrics, and so the Riemannian manifold $M$ itself is changed by the inclusion of $f$.  

\section{Steady-States and Stability Criteria}\label{stability}

In the thermodynamic theory of irreversible processes, a steady-state of order $\kappa$, where $k$ is less than or equal to the dimension $n$ of our thermodynamic space $M$, is a thermodynamic point $x$ with coordinates 
\begin{eqnarray}\label{stab1}
&&U^{1},\ U^{2}\dots U^{\kappa}\quad\qquad{\rm kept\ constant}\quad (\kappa\leq n)\nonumber\\
&&J_{a}=0\quad\qquad\quad\qquad (a=\kappa+1,\cdots n)
\end{eqnarray}
From now on, Latin indices will run from $\kappa +1$ to $n$ and Greek indices will run from $1$ to $n$. Thus (\ref{stab1}) may be written more compactly as 
\begin{equation}\label{stab2}
J_{\mu}dU^{\mu}\mid_{st.state}=0
\end{equation}
\noindent where symbol $d$ denotes the total differential on the submanifold $N$ of solutions of the constraint equations.  

Steady-states may be stable or unstable.  According to the Layapunov stability theory the stability of a state is determined as follows \cite{prigogine}
\begin{eqnarray}\label{stab8}
&{\rm if}&\ \frac{d}{dt}\delta^2 S\mid_{U_0^{\mu}}\ <\ 0\qquad{\rm then\ U_0^{\mu}\ is\ unstable\ for\ t\geq t_0}\nonumber\\
&{\rm if}&\ \frac{d}{dt}\delta^2 S\mid_{U_0^{\mu}}\ >\ 0\qquad{\rm then\ U_0^{\mu}\ is\ asymptotically\ stable \ for\ t\geq t_0}
\end{eqnarray}
\noindent where $S$ denotes the total entropy of the thermodynamic system.  In practice we can modify the value of $\delta^2 S$ by varying a series of control parameters $\lambda$. When these parameters reach a critical value $\lambda_c$, the sign of the inequality (\ref{stab8}) will change, and a steady-state will lose its stability. A steady-state in a configuration with critical control parameters is referred to as a state with marginal stability:
\begin{equation}\label{stab9}
\frac{d}{dt}\delta^2 S\mid_{U_0^{\mu}}(\lambda_c)=0\qquad{\rm for\ t\geq t_0}
\end{equation}

The total entropy $S$ is related to the entropy production $\sigma$ considered in this note via the entropy balance equation
\begin{equation}\label{9a}
\frac{d}{dt} S= \sigma +\Sigma_S
\end{equation}
where $\Sigma_S$ denotes the entropy flux through the boundaries. Eq.~(\ref{9a}) may be varied twice to obtain
\begin{equation}\label{stab10}
\frac{d}{dt}\delta^2 S=\delta^2 \sigma+\delta^2  \Sigma_S
\end{equation}
\noindent 
where the variation of the $\Sigma_S$ term vanishes because we consider homogeneous configurations in which $\Sigma_S$ is constant. In ref. \cite{prigogine}, the authors considered inhomogeneous thermodynamic systems subject to boundary conditions in which all fluxes vanish at infinity. We are instead considering finite homogeneous systems for which the entropy flux through the boundaries, $\Sigma_S$ is constant. Notice that in the particular case $\Sigma_S=0$ the only steady-state is equilibrium. In all other cases our systems do not satisfy their boundary conditions, as they are considering all of spacetime whereas we consider only a region in which the system is homogeneous.  As a result simplifications that occur in ref. \cite{prigogine} which rely upon their choice of boundary conditions do not apply here, but they will not be needed. Thus in our case the stability condition is
\begin{eqnarray}\label{stab11}
&{\rm if}&\ \delta^2 \sigma \mid_{U_0^{\mu}}\ <\ 0\qquad{\rm then\ U_0^{\mu}\ is\ unstable\ for\ t\geq t_0}\nonumber\\
&{\rm if}&\ \delta^2 \sigma \mid_{U_0^{\mu}}\ >\ 0\qquad{\rm then\ U_0^{\mu}\ is\ asymptotically\ stable \ for\ t\geq t_0}
\end{eqnarray}

These conditions may be interpreted geometrically as follows.  Recall that the point $x$ is in an $j$ dimensional submanifold $N\subset M$ of configurations that satisfy a set of physical constraints that relate the thermodynamic forces.  Then $x$ is a steady state of order $\kappa$ if $j=n-\kappa$ and if the derivatives of $\sigma$ with respect to all tangent vectors to $N$ vanish.  Diagonalizing the matrix of second derivatives we may also define the stable directions to be the positive eigenspace, the unstable directions to be the negative eigenspace, and the marginally stable directions to be the null eigenspace.

The steady-state condition (\ref{stab2}) is not canonically defined as the $U$'s at various points $x$ live in different spaces. However using the geometric interpretation of steady-states as critical points of $S$ one arrives at a unique covariant definition for steady-states, and thus a definition that we conjecture will hold beyond the weak-field regime.  The steady states are critical points of $\sigma$ which means that $d\sigma$ vanishes where $d$ is the exterior derivative on the submanifold $N$.  As $\sigma$ is a scalar, the ordinary and covariant derivatives of $\sigma$ are equal and so
\beq 
0=d\sigma=D\sigma=D(U^\mu U^\nu g_{\mu\nu})=2g_{\mu\nu}U^{\nu} DU^\mu
\eeq
where the $f$ term vanishes because $f$ is antisymmetric and $UU$ is symmetric and $Dg$ vanishes because the covariant derivative of the metric always vanishes.  Thus the covariant form of the steady-state condition is just
\beq
g_{\mu\nu}U^{\nu} DU^\mu=0. \label{scov}
\eeq

The stability condition (\ref{stab11}) is already covariant, as $S$ is a scalar and so ordinary derivatives on $S$ are canonically defined.  But for completeness we evaluate the double variation 
\begin{equation}\label{stab12}
\delta^2\sigma=\frac{\partial^2\sigma}{\partial x^a\partial x^b}\delta x^a\delta x^b
\end{equation}
\noindent where
\begin{eqnarray}\label{stab13}
\!\!\!\!\!\!\!\frac{\partial^2\sigma}{\partial x^a\partial x^b}=2g_{\mu\nu}U^{\mu}_{;a}U^{\nu}_{;b}\sp 2f_{\mu\nu;a}U^{\nu}_0U^{\mu}_{;b}\sp 2(g_{\mu\nu}\sp f_{\mu\nu})U_0^{\nu}\Bigl(U^{\mu}_{;a;b}\sp\Gamma^{\kappa}_{ab}U^{\mu}_{;\kappa}\sp U^{\mu}_{;\kappa}\frac{\partial^2x^{\kappa}}{\partial x^a\partial x^b}\Bigr)
\end{eqnarray}
\noindent We have again used the fact that $g$ is covariantly constant.

There are marginally stable states when the determinant of the symmetric matrix $\frac{\partial^2\sigma}{\partial x^a\partial x^b}$ vanishes 
\begin{equation}\label{stab14}
\Bigg\arrowvert (2g_{\mu\nu}\sp f_{\mu\nu})U_0^{\nu}\Bigl( U^{\mu}_{;a;b}\sp \Gamma^{\kappa}_{ab}U^{\mu}_{;\kappa}\sp U^{\mu}_{;\kappa}\frac{\partial^2x^{\kappa}}{\partial x^a\partial x^b}\Bigr)\sp 2g_{\mu\nu}U^{\mu}_{;a}U^{\nu}_{;b}\sp f_{\mu\nu;a}U_0^{\nu}U^{\mu}_{;b}\sp f_{\mu\nu;b}U_0^{\nu}U^{\mu}_{;a}\Bigg\arrowvert_{\lambda=\lambda_c}\!\!\!\!\!\!\!\!\!\!\!\! =0
\end{equation}
We expect that, even beyond the weak-field approximation, Eqs~(\ref{scov}) and (\ref{stab14}) will respectively determine the steady-states $U_0$ and the critical values $\lambda_c$ of control parameter $\lambda$.

\section{Conclusions}\label{conclusions}
We have presented a manifestly covariant and coordinate-independent formulation of the Thermodynamic Field Theory (TFT).  This allows the theory to be extended, for example, to thermodynamic spaces $M$ that cannot be covered by a single coordinate patch corresponding to theories for which different observable quantities are suitable in different regimes. Making use of the Shortest Path Principle, we have demonstrated the validity of the Universal Criterion of Evolution, expressed in a covariant form, and we have shown that this term has a {\it local minimum} at the geodesic line. We referred to this theorem, together with the Shortest Path Principle, as the {\it Minimum Dissipation Principle}. Physically, the Minimum Dissipation Principle affirms that, for time-independent boundary conditions, a thermodynamic system evolves towards a steady-state with the least possible dissipation. 

Of course the test of any proposed formulation is its agreement with experiment.  One physical system for which a large amount of data is already available, and a theoretical underpinning is lacking, is the relaxation of a magnetically confined plasma towards a steady-state in the nonlinear classical and Pfirsch-Schl{\"u}ter regimes or in the nonlinear Banana and Plateau regimes. The metrics in these regimes can be found in \cite{sonnino6}. 

We have also derived the thermodynamic field equations. For the $\alpha-\alpha$ or $\beta-\beta$ processes, these equations reduce to the ones already found in refs \cite{sonnino}-\cite{sonnino3}. The validity of these equations, in the weak-field approximation, have been tested analyzing several thermodynamic systems. So far, we have found that numerical simulations and experiments are in agreement with the theoretical predictions of the TFT. When the skew-symmetric tensor $f_{\mu\nu}$ does not vanish, we have proposed a new set of field equations, which are conjectured to hold even beyond the weak-field regime. The strong field equations are, however, notably simpler than the ones obtained in ref. \cite{sonnino3}. 

The equations that determine the steady-states and the critical values of the control parameters of a generic thermodynamic system have also been presented. These equations are also conjectured to hold beyond the weak-field regime and so they generalize the weak-field equations reported in refs \cite{sonnino1}-\cite{sonnino3}. We note that this extension of linear thermodynamics is quite different from the Prigogine-Glansdorff extension, which is based on hydro- (or plasma-) dynamical stability theory and bifurcation analysis \cite{prigogine}. The relation between the two approaches should be made explicit. This will be the subject of a further publication.

The current approach describes the trajectory followed by a relaxing system, but does not give the time dependence of the relaxation process.  It may be possible to include the time direction in the thermodynamic space with a metric such as $g_{tt}=-\sigma$.  The geodesic evolution in spacetime would then give a firm prediction for the time dependence of the spatial trajectory.  In addition the steady-state condition and the Layapunov stability conditions may be derived from the constraint that particles follow spacetime geodesics.  Of course in order for the system to stop at a steady state somehow dissipation will need to be added to the system, once this is done it may be possible to apply TFT to processes other than relaxation. 

\appendix

\section{Example: Flat Space}\label{flat}
Let $M=\pi$ be a plane. We will compute the map $\exp : T\pi\mapsto \pi$. In this case the Riemannian manifold $(M,g)$ coincides with $(M,L)$ where $L$ is the Onsager matrix. There is a well-defined exponential map from a ball $\mathcal{U}_x$ in $T_x\pi$ to a neighborhood ${\bar {\mathcal U}}_x$ of $x\in \pi$:
\begin{equation}\label{flat1}
{\mathcal U}_x\subset T_{x}\pi=R^2\rightarrow {\bar{\mathcal U}}_x\subset \pi=R^2
\end{equation}
\noindent When the metric is flat, the geodesics $\gamma_V(\tau)$ are straight lines 
\begin{equation}\label{flat2}
x_f^{\mu}=\tau A^{\mu}+x^{\mu}
\end{equation}
\noindent where $A^{\mu}$ is a constant vector and $A\in T_x\pi$, $\gamma_A(t_i)=x$ and $\gamma_A(t_f)=x_f$. Therefore the exp map on $x$ is simply
\begin{equation}\label{flat3}
\exp_x\tau A=\tau A+x
\end{equation} 
\noindent that is, the exponential map $\exp_p V$ (with $p\in\pi$ and $V\in T_p\pi$) simply performs a translation:
\begin{equation}\label{flat4}
\exp_pV=p+V
\end{equation} 
\noindent Let us now consider the case $U\in\mathcal{U}_x\subset T_0M=R^2$ where $0\in\pi$ denotes the origin, corresponding to the equilibrium state. In this case $U$ points from $x$ to the origin and, using the canonical identification of the plane $\pi$ and its tangent fibers, which are planes of the same dimension, we find the relation $x=-U$ (see Eq.~(\ref{equil1})). If the coordinates of $x$ are $x=(a,b)$, the coordinates of $U$ are then $U=(-a,-b)$ and, from Eq.~(\ref{flat4}), we have 
\begin{equation}\label{flat5}
\exp_pU=p+U=(a,b)+(-a,-b)=0
\end{equation} 
\noindent This implies that the exponential map sends the tangent vector $U$ at $p$ to the equilibrium state, which is the definition of $U$ in Eq.~(\ref{vt2}).

The entropy production of a system at the point $x$ is just its length squared
\beq
\sigma=|U|^2=|-x|^2=|x|^2
\eeq

\section{Example: The Two-Sphere}\label{sphere}

Let $M$ be the unit two-sphere $S^2$ with the usual rotation-invariant metric
\beq
ds^2=d\th^2+\sin^2\th d\phi^2.
\eeq
We will use both spherical and Cartesian coordinate systems, which are related by
\beq
x=\sin\th\ \cos\phi,\quad y=\sin\th\ \sin\phi,\quad z=\cos\th.
\eeq
The geodesics on the 2-sphere are the great circles, which are the equators corresponding to various choices of north pole.  We will now fix the north pole to be $\th=0$, and so every geodesic $\gamma$ will intersect the equator $\th=\pi/2$ at two points $\phi=-\delta$ and $\phi=\pi-\delta$.  Note that the equator itself is the geodesic that intersects itself an infinite number of times, to apply the following derivation to the equator itself it would be necessarily to deform it by a small rotation and then take that rotation to zero at the end.  However the final result will make sense for every geodesic, including the equator.

The map $exp(\vec{v})$ is easy to calculate in the case in which $\gamma$ is the equator, it just increases $\phi$ by the norm $v$ of the vector $\vec{v}$.  To evaluate $exp$ on an arbitrary point $(\th_0,\phi_0)$ with an arbitrary tangent vector $\vec{v}$, which yields an arbitrary geodesic $\gamma$, it will be simplest to rotate the coordinate system so that $\gamma$ becomes the equator, then $exp$ may be evaluated and the coordinate system may be rotated back to yield the final answer.  

The rotation of the coordinate system will proceed in two steps.  First we use $\delta$ to define a rotated set of coordinates
\beq
\tilde{x}=\sin\th\ \cos(\phi+\delta),\quad \tilde{y}=\sin\th\ \sin(\phi+\delta),\quad \tilde{z}=z=\cos\th.
\eeq
Then we rotate about the $\tilde{y}$-axis until $\gamma$ is the equator, which we parametrize with the azimuthal variable $\tau$
\beq \label{rot}
\left(\begin{array}{c}
\tilde{x}\\\tilde{y}\\\tilde{z}\end{array}\right)=
\left(\begin{array}{ccc}
1&0&0\\0&\cos\a&-\sin\a\\0&\sin\a&\cos\a\end{array}\right)
\left(\begin{array}{c}
\cos\tau\\\sin\tau\\0\end{array}\right).
\eeq 
In coordinates the rotation (\ref{rot}) is
\beq
\sin\th\ \cos(\phi+\delta)=\cos\tau,\quad \sin\th\ \sin(\phi+\delta)=\cos\a\ \sin\tau,\quad \cos\th=\sin\a\ \sin\tau. \label{rotco}
\eeq

The tangent vector $\vec{v}$ will be decomposed in polar coordinates $(v,\b)$ where $v$ is the norm of $\vec{v}$ and $\beta$ is the angle between $\vec{v}$ and the azimuthal direction $\phi$.  In this section we will calculate $exp_{(\theta_0,\phi_0)}\vec{v}$.  First we will find expressions satisfied by the unknown angles $\tau$ and $\alpha$ in terms of the given data $\theta_0,\ \phi_0, v, \b$.  These will allow us to calculate the $z$ component $z\p$ of the function $exp$, which is particularly simple as it is independent of $\delta$.  Then we will rotate the coordinates to construct the other components $x\p$ and $y\p$.

$\vec{v}$ is a tangent vector to the geodesic $\gamma$ and so the rotation (\ref{rot}) relates $\vec{v}$ to the equator's tangent vector $\partial/\partial\tau$.  For example the last expression in (\ref{rotco}) yields
\beq
\frac{\partial}{\partial\tau}\cos\th=\frac{\partial}{\partial\tau}\sin\alpha\ \sin\tau \label{deriv}
\eeq
where the derivative is evaluated at $(\th_0,\phi_0)$.  The left hand side may be evaluated using the chain rule, where we note that the derivative of $\th$ with respect to $\tau$ along the geodesic is just the theta component $\sin\b$ of the unit tangent vector $\vec{v}/v$.  The right hand side is evaluated by noting that $\alpha$ is the angle by which the entire geodesic $\gamma$ is rotated, and so it is independent of $\tau$.  Thus (\ref{deriv}) reduces to
\beq
\sin\beta\ \sin\th_0=\sin\a\ \cos\tau.
\eeq
The dependence on the unknown angle $\a$ may be removed using the last term in Eq.~(\ref{rot}).  We then divide both sides of the resulting equation by $\cos\th_0$ to obtain
\beq
\cot\tau=\sin\beta\ \tan\th_0. \label{tau}
\eeq
Thus we have determined the unknown angle $\tau$ in terms of the initial point and tangent vector.  While in principle we could also find $\alpha$, it will suffice to recall the last term from (\ref{rotco})
\beq
\sin\a=\cos\th_0\ \csc\tau. \label{sin}
\eeq

In our new coordinates the $exp$ operation corresponds to a simple shift of the equator's longitudinal coordinate $\tau$
\beq
\tau\mapsto\tau+v.
\eeq
Finally we need to reexpress this shift in terms of the old coordinates.  In fact it will suffice to find the corresponding coordinate $z\p$.  Using the rotation (\ref{rot}) at the new coordinate $\tau+v$
\beq
\left(\begin{array}{c}
x\p\\y\p\\z\p\end{array}\right)=
\left(\begin{array}{ccc}
1&0&0\\0&\cos\a&-\sin\a\\0&\sin\a&\cos\a\end{array}\right)
\left(\begin{array}{c}
\cos(\tau+v)\\\sin(\tau+v)\\0\end{array}\right).
\eeq
we find
\beq
z\p=\sin\a\ \sin(\tau+v).
\eeq
This may be expanded using (\ref{sin}) and the trigonometric identity for the sine of a sum to yield
\beq
z\p=\cos\th_0\ \csc\tau (\sin\tau\ \cos v+\cos\tau\ \sin v)=\cos\th_0(\cos v+\cot\tau\ \sin v).
\eeq
Finally we may use (\ref{tau}) to obtain
\beq
z\p=\cos\th_0\ \cos v+\sin\b\ \sin\ \th_0\ \sin v. \label{z}
\eeq

To find the corresponding formula for $x\p$ we may rotate the coordinates by $\pi/2$, yielding a new $z$ coordinate, $\tilde{z}$, equal to the old $x$ coordinate
\beq
\tilde{z}=\cos\tilde{\th}=\sin\th\ \cos\phi=x.\label{rotgrande}
\eeq
This rotation is an isometry so $v$ is unchanged, however the angle $\b$ is effected.  The rotated value of $\beta$ may be found similarly to (\ref{deriv})
\beq
\frac{\partial}{\partial\tau}\cos\tilde{\theta}=\sin\tilde{\b}\ \sin\tilde\th=\frac{\partial}{\partial\tau}(\sin\th\ \cos\phi)=\sin\b\ \cos\th\ \cos\phi-\cos\b\ \sin\phi. \label{bnuovo}
\eeq
In the last term of the last expression of (\ref{bnuovo}) we have used
\beq
\cos\b=g_{\phi\phi}^{1/2}\frac{\partial\phi}{\partial\tau},\quad g_{\phi\phi}=\sin^2\th.
\eeq

We may now use (\ref{rotgrande}) and (\ref{deriv}) to rotate our equation (\ref{z}) for the $z$ component of the $exp$ function to obtain the $x$ component
\beq
x\p=\sin\th_0\ \cos\phi_0\ \cos v+(\sin\b\ \cos\th_0\ \cos\phi_0-\cos\b\ \sin\phi_0)\sin v. \label{x}
\eeq
And at last we may rotate $\phi$ to obtain the $y$ component
\beq
y\p=\sin\th_0\ \sin\phi_0\ \cos v+(\sin\b\ \sin\th_0\ \cos\phi_0+\cos\b\ \cos\phi_0)\sin v. \label{y}
\eeq
Putting together Eqs.~(\ref{z},\ref{x},\ref{y}) we obtain
\beq
exp_{(\theta_0,\phi_0)}(v,\beta)=(x\p,y\p,z\p).
\eeq

Taking the north pole to be equilibrium, the thermodynamic force at any point is calculated using the geodesic that runs straight north from any point to the north pole
\beq
U=(v=\theta,\beta=0).
\eeq
This geodesic is uniquely determined for all points except for the north pole $\theta=\pi$, for which there are an infinite number of northward geodesics and the coordinate $\beta$ is ill-defined.  The entropy production is then simply
\beq
\sigma=\langle U, U\rangle = \theta^2
\eeq
Note that at the south pole, where the geodesic is not unique, $\sigma$ is independent of the geodesic chosen and so continues to be well defined.

\section*{Acknowledgments}

\noindent We acknowledge our indebtedness to Prof. C.M. Becchi of the University of Genova - Physics Department and to Prof. E. Tirapegui of the University of Chile - Physics Department, for reading this manuscript and making helpful suggestions. 

\noindent One of us (GS)  would like to express his sincere gratitude to his hierarchy at the European Commission: Prof. A. Mitsos, Dr P. Fernandez Ruiz and Dr E. Rille. He is also very grateful to Dr U. Finzi and Dr M. Cosyns, of the European Commission, for their continuing encouragement.

\noindent GS would like to thank the members of the EURATOM Belgian State Fusion Association, in particular Dr D. Carati, Dr B. Knaepen and Dr B. Weyssow. 

\noindent The work of JE is partially supported by IISN - Belgium (convention 4.4505.86), by the ``Interuniversity Attraction Poles Programme -- Belgian Science Policy'' and by the European Commission RTN program HPRN-CT-00131, in which he is associated to K. U. Leuven.


\begin{thebibliography}{alpha}
\bibitem{sonnino} Sonnino G. {\it Il Nuovo Cimento}, {\bf 115 B}, (2000), 1057.
\bibitem{sonnino1} Sonnino G. 2001 {\it Thermodynamic Field Theory (An Approach to Thermodynamics of Irreversible Processes)}, proceedings of the {\it 9th International Workshop on Instabilities and 
Nonequilibrium Structures}, Vi$\tilde{\mathrm n}$a del Mar (Chile). {\it Kluwer Academic Publishers}, {\it Nonlinear Phenomena and Complex Systems Editions}, {\bf 9}, (2004) 291.
\bibitem{sonnino2} Sonnino G. 2002 {\it A Field Theory Approach to Thermodynamics of Irreversible Processes}, (Th{\` e}se d'Habilitation {\`a} Diriger des Recherches - H.D.R.) 
- Institut Non Lin{\`e}aire de Nice (France).
\bibitem{sonnino3} Sonnino G. {\it  Il Nuovo Cimento}, {\bf 118 B}, (2003) 1175. 
\bibitem{sonnino4} Sonnino G. {\it Int. J. of Quantum Chemistry} {\bf 98} (2004) 191.
\bibitem{sonnino5} Sonnino G. 2005 {\it A Field Theory Approach to Transport Processes in Confined Plasmas: the Nonlinear Classical and Pfirsch-Schl{\"{u}}ter Regimes and Transport Equations in the Nonlinear Banana and Plateau Regimes} (submitted to publication in the review {\it I.A.E.A. Nuclear Fusion}). 
\bibitem{peeters} Peeters Ph. and Sonnino G. {\it Il Nuovo Cimento} {\bf 115 B} (2000) 1083.
\bibitem{sonnino6} Sonnino G. {\it Nuovo Cimento} {\bf 118 B} (2003) 1155. 
\bibitem{hall} Sonnino G. and Peeters Ph. {\it Chaos} {\bf 14} (2004) 910.
\bibitem{prigogine1} Prigogine I. 1954 {\it Thermodynamics of Irreversible processes}, (John Wiley \& Sons). 
\bibitem{prigogine} Glansdorff P. and Prigogine I. 1971 {\it Thermodynamics of Structures, Stability and Fluctuations}, (John Wiley \& Sons).
\bibitem{katok} Katok A. and Hasselblatt B. 1995 {\it Introduction to the Modern Theory of Dynamical Systems}, (Cambridge University Press).
\bibitem{fermi} Fermi F.{\it Atti R. Accad. Lincei, Cl. Sc. Fis. Mat. Nat.} {\bf 31} (1922) 21. 
\bibitem{weinberg} Weinberg S. 1972 {\it Gravitation and Cosmology: Principles and Applications of the General Theory of Relativity}, (John Wiley \& Sons).
\bibitem{frankel} Frankel T.  1997 {\it The Geometry of Physics - An Introduction}, (Cambridge University Press).
\bibitem{synge} Synge J.L. and Schild A. 1949 {\it Tensor Calculus}, (Dover Publications, Inc. New York).
\bibitem{eisenhart} Eisenhart L.P. 1997 { \it Riemannian Geometry}, (Eight printing, for Princeton Landmarks in Mathematics and Physics series - Princeton University Press). 
\bibitem{lovelock} Lovelock D. and Rund H.  1989 {\it Tensors, Differential Forms and Variational Principles}, (Dover Publications, Inc., New York). 
\bibitem{york} York J.W. Jr. {\it Found. Phys.}, {\bf 16}, (1986) 249.
\bibitem{lefever} Horsthemke H. and Lefever R.  1984 {\it Noise-Induced Transitions. Theory and Applications in Physics, Chemistry and Biology}, (Springer-Verlag). 
\bibitem{string} Green M.B., Schwarz J.H. and Witten E. 1987 {\it Superstring Theory. Volume 1 Introduction}, (Cambridge University Press). 
\end{thebibliography}
\end{document}